\documentclass[aps, twocolumn,superscriptaddress,floatfix,showpacs,amsmath,longbibliography,nofootinbib,amssymb]{revtex4-1}
\usepackage[colorlinks=true,citecolor=blue,linkcolor=blue]{hyperref}
\usepackage[usenames]{color}
\usepackage{subfigure}
\usepackage{dcolumn}
\usepackage{mathrsfs}
\usepackage{bm}
\usepackage{amsmath,amssymb,epsfig,float,graphics}
\usepackage{appendix}

\begin{document}
\baselineskip=0.45 cm

\title{Using nanokelvin quantum thermometry to detect timelike Unruh effect in a Bose-Einstein condensate}

\author{Zehua Tian}
\email{tianzh@ustc.edu.cn}
\affiliation{CAS Key Laboratory of Microscale Magnetic Resonance and School of Physical Sciences, University of Science and Technology of China, Hefei 230026, China}
\affiliation{CAS Center for Excellence in Quantum Information and Quantum Physics, University of Science and Technology of China, Hefei 230026, China}

\author{Jiliang Jing}
\affiliation{Department of Physics, Key Laboratory of Low Dimensional Quantum Structures and Quantum Control of Ministry of Education, and Synergetic Innovation Center for Quantum Effects and Applications, Hunan Normal University, Changsha, Hunan 410081, P. R. China}


\begin{abstract}
It is found that the Unruh effect can not only arise out of the entanglement between two sets of modes spanning the left and right Rindler wedges, but also 
between modes spanning the future and past light cones. Furthermore, an inertial Unruh-DeWitt detector along a spacetime trajectory in one of these cones may 
exhibit the same thermal response to the vacuum as that of an accelerated detector confined in the Rindler wedge. 
This feature thus could be an alternative candidate to verify the ``Unruh effect", termed as the timelike Unruh effect correspondingly. In this paper we propose to detect the timelike Unruh effect by using an impurity immersed in a Bose-Einstein condensate (BEC). The impurity acts as the detector which interacts with the density fluctuations in the condensate, working as an effective quantum field. Following the paradigm of the emerging field of \emph{quantum thermometry}, we combine quantum parameter estimation theory with the theory of open quantum systems to realize a nondemolition Unruh temperature measurement in the nanokelvin ($\mathrm{nK}$) regime. Our results demonstrate that the timelike Unruh effect can be probed using a stationary two-level impurity with time-dependent energy gap immersed in a BEC within current technologies.

\end{abstract}

\baselineskip=0.45 cm
\maketitle
\newpage

\section{Introduction} 
The Unruh effect  \cite{PhysRevD.14.870} is a conceptually subtle quantum field theory result in relativistic framework. It plays a crucial role in our understanding 
that vacuum fluctuations and the particle content of a field theory are observer-dependent. It predicts that uniformly accelerating observers perceive the 
quantum field vacuum defined by inertial observers as a thermal state, rather than a zero-particle state. 
Since, to produce an experimentally appreciable Unruh temperature, prohibitively large accelerations  have to be required (e.g., about $1\,\mathrm{Kelvin}$ even for accelerations as high as $10^{20}\,\mathrm{m/s^2}$), the direct experimental confirmation of the Unruh effect until now still remains elusive.

A uniformly accelerating observer is conveniently described as a stationary observer in Rindler coordinates \cite{PhysRev.119.2082, birrell1984quantum}. 
From the perspective of the accelerating observer, the Minkowski vacuum state of a quantum field, e.g., a massless scalar field, defined by the inertial observers can be written as a spacelike entangled state between two sets of modes, respectively, spanning the left and right Rindler wedges \cite{RevModPhys.80.787}. 
Since the accelerating observer is confined 
to just one of these wedges, the thermalized vacuum is obtained (i.e., Unruh effect arises as the result of the spacelike entanglement) when tracing out the unobserved modes. To detect the Unruh effect, many experimental scenarios have been designed in the \emph{analogue gravity} \cite{Analogue-Gravity} regime, including various
physical systems ranging from ultracold atoms systems \cite{PhysRevLett.101.110402, PhysRevA.103.013301, PhysRevResearch.2.042009, PhysRevD.106.L061701, PhysRevLett.125.213603, Unruh-effect1, Unruh-BEC, bunney2023sound} to a graphene nanosheet  system \cite{PhysRevLett.126.117401}.
In this regard, let us note that quantum simulation of Unruh effect has been reported through the BEC system \cite{Unruh-effect1} and NMR system \cite{Unruh-effect-NMR} experimentally. These simulations rely on functional equivalence (i.e., simulating two-mode squeezed mechanics), while one significant step forward is to embody the essential of Unruh effect---\emph{acceleration-induced} particle creation---in the simulation. Furthermore, other scenarios to enhance 
the detection of the acceleration-induced emission has been successfully theoretically put forward \cite{PhysRevLett.91.243004, PhysRevLett.107.131301, PhysRevLett.125.241301, PhysRevD.106.045011, PhysRevLett.129.111303}. However, the intractable, but required, relativistic motion is still the main obstacle to verifying the Unruh effect in practice.

\begin{figure}
\centering
\includegraphics[width=0.32\textwidth]{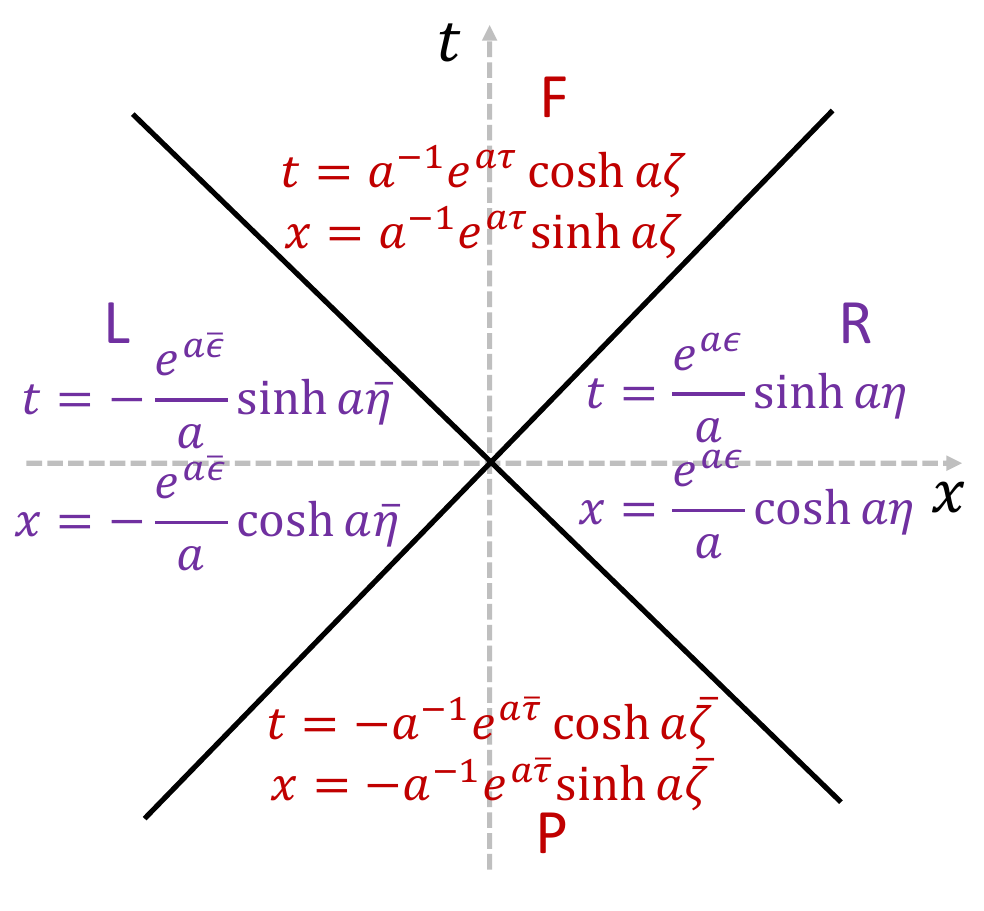}
\caption{A spacetime diagram divided into four quadrants: the future (F) and past (P) light cones, and the left (L) and right (R) Rindler wedges. The transform between 
the usual Minkowski coordinate $(t, x)$ and that of the four regimes are shown. From the perspective of an accelerating observer, the Minkowski vacuum of massless quantum field can be expanded in terms of modes confined in R and L or in terms of modes restricted to F and P. Therefore, if the observer in one of these quadrants (e.g., the F light cone), tracing over the unobserved modes (e.g., in the P light cone) leads to the (timelike) Unruh effect.}\label{fig1}
\end{figure}

Recently, it has been shown that the Minkowski vacuum state of a quantum field could also be written down similarly as entangled states between modes spanning
in the future and past light cones \cite{PhysRevLett.106.110404, PhysRevD.96.083531, PhysRevD.103.125005}. The Unruh effect may also arise 
as a result of this timelike entanglement if an observer (or detector interacting with the field) is confined only in one of these cones. In this case, a detector in the future/past light cone with world line $(\tau, 0)$ (see Fig. \ref{fig1}) corresponds to a detector with energy gap scaled with $1/at$ in the laboratory frame (see more details below). Therefore, one can in principle detect this timelike Unruh effect by considering a stationary detector with a special form of time-dependent energy gap \cite{PhysRevLett.106.110404}, while without involving any real relativistic motion. Note that extraction of timelike entanglement from the quantum vacuum \cite{PhysRevA.85.012306} has been investigated with this model. Besides, Berry phase from the entanglement of future and past light cones has been proposed 
to detect this timelike Unruh effect \cite{PhysRevLett.129.160401}. However, a concrete experimentally feasible scenario to estimate the timelike Unruh effect is still lacking. 

In this paper, we aim at resolving this issue by proposing to detect the timelike Unruh effect with an experimentally accessible platform consisting of a BEC
and an immersed impurity \cite{PhysRevLett.94.040404, PhysRevLett.91.240407}. In our scenario, the density fluctuations in the BEC are modeled as the quantum field.
The impurity, analogously dipole coupled to the density fluctuations, acts as an Unruh-DeWitt detector, and its energy gap can be designed as a time-dependent one by an external electromagnetic field (see more details below). We treat the impurity as an open system and derive its dynamical evolution by tracing over the degree of freedom of the analogous quantum field. Adopting tools from the theory of quantum parameter estimation, we show that our proposed scheme achieves experimentally 
accessible precision for the Unruh temperature detection in the BEC system within current technologies. 

The outline of this paper is as follows. In Sec. \ref{section2} we review some basic concepts and physics of the Unruh effect from the perspective of 
quantum field theory and Unruh-DeWitt detector approach. In Sec. \ref{section3} we propose a concrete  scenario to realize the detection of the timelike Unruh effect using the BEC platform. Experimental feasibility of the relevant analysis within the current technologies of BEC is discussed in Sec. \ref{section4}.  Finally, discussion and summary of the main results are present in Sec. \ref{section5}.


\section{Timelike Unruh effect} \label{section2}
In this section we will simply review the timelike Unruh effect from the perspective of quantum field vacuum state and Unruh-DeWitt detector response. For more details, we refer the reader to Refs. \cite{RevModPhys.80.787, PhysRevLett.106.110404}.

\subsection{Quantum field vacuum state}
Let us begin with the spacetime broken into quadrants (\text{F, P, R, L}) shown in Fig. \ref{fig1}.
The corresponding four coordinate systems can be used to define a set of field modes, complete in each region. To derive the 
Unruh effect, we here, for simplification, consider a two dimensional massless  scalar field case for the example.  Because of
the conformal invariance of the massless wave equation in two dimensions, all the Klein-Gordon equations in the four coordinate systems
are the same as that of two-dimensional Minkowski case. They are given by \cite{RevModPhys.80.787},
\begin{eqnarray}\label{KG-equation}
\bigg(\frac{\partial^2}{\partial\tau^2}-\frac{\partial^2}{\partial\zeta^2}\bigg)_\text{F}\phi&=&0,~~~~\bigg(\frac{\partial^2}{\partial\eta^2}-\frac{\partial^2}{\partial\epsilon^2}\bigg)_\text{R}\phi=0,
\\
\bigg(\frac{\partial^2}{\partial\bar{\tau}^2}-\frac{\partial^2}{\partial\bar{\zeta}^2}\bigg)_\text{P}\phi&=&0,~~~~\bigg(\frac{\partial^2}{\partial\bar{\eta}^2}-\frac{\partial^2}{\partial\bar{\epsilon}^2}\bigg)_\text{L}\phi=0,
\end{eqnarray}
where the subscripts $(\text{F}, \text{P}, \text{R}, \text{L})$ denote the regions. Correspondingly, the set of field modes defined in the respective quadrants are given by
\begin{eqnarray}
\phi^\text{R}_\omega(\chi)&=&(4\pi\omega)^{-1/2}e^{-i\omega\chi},~\phi^\text{L}_\omega(\bar{\chi})=(4\pi\omega)^{-1/2}e^{-i\omega\bar{\chi}},
\\
\phi^\text{F}_\omega(\nu)&=&(4\pi\omega)^{-1/2}e^{-i\omega\nu},~\phi^\text{P}_\omega(\bar{\nu})=(4\pi\omega)^{-1/2}e^{-i\omega\bar{\nu}},
\end{eqnarray}
where $\chi=\eta+\epsilon$, $\bar{\chi}=-\bar{\eta}-\bar{\epsilon}$, $\nu=\tau+\zeta$, and $\bar{\nu}=-\bar{\tau}-\bar{\zeta}$ denote the ``light-cone" in the four coordinate systems above. 

One can quantize the field and define the field vacuum and field creation/annihilation operators in the respective quadrants. For example, one can define the Rindler vacuum as $\hat{a}^\text{R}_{\omega}|0_\text{R}\rangle=0$ with $\hat{a}^\text{R}_{\omega}$ being the annihilation operator for a right Rindler particle, corresponding 
to the solution $\phi^\text{R}_\omega(\chi)$. Here $\omega$ here is the particle frequency. The particle number state is given by $\frac{(\hat{a}^{\text{R}\dagger}_{\omega})^n}{n!}|0_\text{R}\rangle=|n_\omega,\text{R}\rangle$ with $\hat{a}^{\text{R}\dagger}_{\omega}$ being the creation operator.

From the perspective of an accelerated 
observer, the vacuum of quantum massless scalar field, $|0_\text{M}\rangle$, defined by an inertial observer in the Minkowski spacetime can be rewritten as an entangled state between two
sets of modes, respectively, spanning the right and left Rindler wedges \cite{PhysRevD.14.870, RevModPhys.80.787}:
 \begin{eqnarray}\label{Vacuum-Rindler1}
 |0_\text{M}\rangle=\prod_{i=1}\bigg(C_i\sum_{n_i=0}^\infty\,e^{-\pi\,n_i\omega_i/a}|n_i, \text{R}\rangle\otimes|n_i, \text{L}\rangle\bigg),
 \end{eqnarray}
where $C_i=\sqrt{1-e^{-2\pi\omega_i/a}}$, and $n_i$ and $\omega_i$ respectively denote the Rindler particle number and energy in corresponding 
regions, $\text{R}$ and $\text{L}$. However, since the R region and L region are causally disconnected (spacelike),  the uniformly accelerated observer can only
access one set 
of Rindler modes. The tracing out of the unobserved modes (e.g., the left Rindler modes) leads to the prediction that such an accelerated observer sees a thermalized vacuum, i.e.,
\begin{eqnarray}\label{spacelike-thermal}
\hat{\rho}_\text{R}=\prod_{i=1}\bigg(C^2_i\sum_{n_i=0}^\infty\,e^{-2\pi\,n_i\omega_i/a}|n_i, \text{R}\rangle\langle\,n_i, \text{R}|\bigg).
\end{eqnarray}
Note that this is density matrix for the system of free bosons with temperature $T=a/2\pi$. It means that the Minkowski vacuum state $|0_\text{M}\rangle$ of quantum field is viewed as a thermal state from the perspective of the accelerated observer, known as the Unruh effect. 

Modes in $R$ is independent from modes in $L$, and modes in $F$ is independent from modes in $P$, but modes in F/P are not independent 
of the modes in R/L \cite{PhysRevLett.106.110404}. Actually, $\phi^\text{F}_\omega(\nu)$ is the same solution as $\phi^\text{R}_\omega(\chi)$, extended from $\text{R}$
into $\text{F}$, which is pointed out in Refs. \cite{RevModPhys.80.787, PhysRevLett.106.110404}. This is also hold for $\phi^\text{P}_\omega(\nu)$ and $\phi^\text{L}_\omega(\chi)$. Therefore, the demonstration of F-P entanglement of the Minkowski vacuum is exactly the same as the standard demonstration of R-L entanglement shown in Eq. \eqref{Vacuum-Rindler1}, with a change of labels $\text{R}\rightarrow\text{F}$ and $\text{L}\rightarrow\text{P}$. That is to say, the Minkowski vacuum restricted to $\text{F}-\text{P}$ takes a symmetrical form when expressed in terms of the ``conformal modes"
$\phi^\text{F}_\omega$ and $\phi^\text{P}_\omega$:
 \begin{eqnarray}\label{Vacuum-Rindler2}
 |0_\text{M}\rangle=\prod_{i=1}\bigg(C_i\sum_{n_i=0}^\infty\,e^{-\pi\,n_i\omega_i/a}|n_i, \text{F}\rangle\otimes|n_i, \text{P}\rangle\bigg).
 \end{eqnarray}
Analogously, the state of the field in the future (or the past) alone is a ``thermal" state of the $\phi^\text{F}_\omega$ ($\phi^\text{P}_\omega$)-modes, given by
\begin{eqnarray}\label{timelike-thermal}
\hat{\rho}_\text{F}=\prod_{i=1}\bigg(C^2_i\sum_{n_i=0}^\infty\,e^{-2\pi\,n_i\omega_i/a}|n_i, \text{F}\rangle\langle\,n_i, \text{F}|\bigg).
\end{eqnarray}
Note that this thermal phenomenon in \eqref{timelike-thermal} arises out of the timelike entanglement between modes of F and P light cones, so 
it is also called as the timelike Unruh effect \cite{PhysRevLett.106.110404, PhysRevA.85.012306}.

\subsection{Unruh-DeWitt detector response function}
To detect the timelike Unruh effect, one can consider an Unruh-DeWitt detector evolving in the F light cone with world line $x=y=z=0, t=a^{-1}e^{a\tau}$.
Along this world line the Schr\"odinger equation in the conformal time $\tau$ reads $i\partial\psi/\partial\tau=H\psi$, and the eigenvalues of $H$ is assumed to
be a constant gap $\omega_0$. In Minkowski coordinates or 
laboratory framework, this Schr\"odinger equation can be rewritten as
\begin{eqnarray}
i\frac{\partial\psi}{\partial\,t}=\frac{H}{at}\psi.
\end{eqnarray}
The $1/at$ factor is due to the change of variables to Minkowski time. This means that in the laboratory a detector 
with energy gap (denoted by the Hamiltonian $H$) scaled with $1/at$ corresponds to a detector with the fixed Hamiltonian $H$ on the  $x=y=z=0, t=a^{-1}e^{a\tau}$ world line \cite{PhysRevLett.106.110404}.

We can assume the interaction between the detector and the field to be the standard Unruh-DeWitt term. Specifically, for a two-level detector it takes the form 
$H_I=\lambda\hat{\phi}(x(t))\sigma_x$. Therefore, consider the full Hamiltonian in the conformal time framework, we have $i\partial\psi/\partial\tau=(H+e^{a\tau}H_I)\psi$. 
The interaction term acquires the exponential factor due to the change of variables to conformal time. 
For this case, the detector response function is found to be 
\begin{eqnarray}\label{RF}
{\cal G}(\omega_0)=\int^\infty_{-\infty}d\tau\int^\infty_{-\infty}d\tau^\prime\,e^{-i\omega_0(\tau-\tau^\prime)}e^{a(\tau+\tau^\prime)}D^+(\tau, \tau^\prime),
\end{eqnarray}
where $D^+(\tau, \tau^\prime)=\langle0_\text{M}|\phi(\tau)\phi(\tau^\prime)|0_\text{M}\rangle$ denotes the Wightman function of the field. The 
Wightman function along the inertial trajectory  $x=y=z=0, t=a^{-1}e^{a\tau}$ can be calculated to take the form \cite{PhysRevLett.106.110404, PhysRevA.85.012306}
\begin{eqnarray}\label{WF1}
D^+(\tau, \tau^\prime)=-\frac{a^2e^{-a(\tau+\tau^\prime)}}{16\pi^2\sinh^2\big[\frac{a}{2}(\tau-\tau^\prime)-i\varepsilon\big]},
\end{eqnarray}
while it along a uniformly accelerated trajectory $t=a^{-1}\sinh(a\eta), x=a^{-1}\cosh(a\eta), y=z=0$ in the R region 
takes the form
\begin{eqnarray}\label{WF2}
D^+(\eta, \eta^\prime)=-\frac{a^2}{16\pi^2\sinh^2\big[\frac{a}{2}(\eta-\eta^\prime)-i\varepsilon\big]}.
\end{eqnarray}
With the Wightman functions in Eqs. \eqref{WF1} and \eqref{WF2}, we can find the response function integral for the inertial detector in F region is formally 
identical to that for the accelerated detector in the R region. 
Furthermore, through the standard evaluation of the response function 
integral \cite{birrell1984quantum}, this leads us to a thermal response function at temperature $T=\frac{a}{2\pi}$, called as the Unruh temperature.
Note that the accelerated detector in R region can demonstrate the Unruh effect in terms of its response function. However,
the inertial detector in F region may also demonstrate the Unruh effect, while without involving any acceleration motion. Instead of that, it requires 
its energy gap scaled with $1/at$ in the laboratory \cite{PhysRevLett.106.110404, PhysRevA.85.012306}.

\section{Detect the timelike Unruh effect in a BEC} \label{section3}
We will in the following 
observe the timelike Unruh effect with an impurity immersed into a BEC, and adopt tools from the theory of quantum parameter estimation to
estimate the Unruh temperature.

\subsection{Dynamics of the impurity detector in a BEC}
To demonstrate the timelike Unruh effect with the Unruh-DeWitt detector model in a BEC, let us begin with the standard one-dimensional Gross-Pitaevskii equation \cite{gross1961structure, gross1963hydrodynamics, pitaevskii1961vortex}
\begin{eqnarray}
i\partial_t\Psi=\bigg[-\frac{1}{2m}\nabla^2+V_\text{ext}+g|\Psi|^2\bigg]\Psi,
\end{eqnarray}
where $\Psi$ is the condensate wave function, $V_\text{ext}$ is the externally imposed trapping potential, and $g$ denotes the  two-body contact interaction coupling.
In the second-quantized formalism, one can decompose the field operator for the dilute Bose gas as $\hat{\Psi}=\Psi_0(1+\hat{\phi})$ with $\Psi_0=\sqrt{\rho_0}e^{i\theta_0}$, where $|\Psi_0(x)|^2=\rho_0\simeq\text{const}$ represents the condensate density, and $\hat{\phi}$ describes the perturbations (excitations) on
 the top of the condensate. Within the Bogoliubov theory \cite{pethick2008bose, pitaevskiui2016bose}, density fluctuations in Heisenberg representation can 
 be written in the form of  $\hat{\phi}$ as 
 \begin{eqnarray}\label{density-fluctuations}
 \nonumber
 \delta\hat{\rho}(t,x)&\simeq&\rho_0(\hat{\phi}+\hat{\phi}^\dagger)=\sqrt{\rho_0}\int\frac{dk}{2\pi}(u_k+v_k)
 \\
 &&\times[\hat{b}_ke^{-\omega_kt+ikx}+\hat{b}^\dagger_ke^{i\omega_kt-ikx}],
 \end{eqnarray}
which closely resembles the quantized scalar field in terms of bosonic operators $\hat{b}_k\,(\hat{b}^\dagger_k)$ satisfying the usual Bose commutation rules 
$[\hat{b}_k, \hat{b}^\dagger_k]=2\pi\delta(k-k^\prime)$. In the laboratory frame where the condensate is at rest, the frequency of quasiparticle reads $\omega_k=c_0k\sqrt{1+(\xi_0k/2)^2}$
with $c_0$ and $\xi_0$ being the speed of sound and the healing length, respectively. Note that $u_k$ and $v_k$ are Bogoliubov parameters
satisfying $(u_k+v_k)^2=E_k/\omega_k$ with $E_k=k^2/2m$.

We consider an impurity as the simulator of the Unruh-DeWitt detector, which is immersed into the condensate and collisionally coupled to the Bose gas. The impurity is assumed 
to be illuminated by a monochromatic external electromagnetic field at the frequency close to resonance with the impurity's internal level transition, with a time-dependent 
Rabi frequency. The effective Hamiltonian of the whole system, the impurity plus the density fluctuations, is given by (see Supplemental Material \cite{Supplemental-Material} for details)
\begin{eqnarray} \label{Interaction-Hamiltonian}
H(t)&=\frac{\omega_0(t)}{2}\sigma_z+g_-\sigma_x\delta\hat{\rho}(x_A, t),
\end{eqnarray}
where $\omega_0(t)$ is the time-dependent Rabi frequency which can be experimentally controlled, and $g_-$ is the coupling parameter. Therefore, the dynamics of
this system yields $i\partial_t|\Psi\rangle=H(t)|\Psi\rangle=[\frac{\omega_0(t)}{2}\sigma_z+g_-\sigma_x\delta\hat{\rho}(x_A, t)]|\Psi\rangle$. If we introduce a new time parameter $\tau$ satisfying $t=c_0a^{-1}e^{a\tau/c_0}$, then in this time frame one can find 
\begin{eqnarray}\label{Equation-sysmtem}
i\partial_\tau|\Psi\rangle=\big[\frac{\omega_0}{2}\sigma_z+e^{a\tau/c_0}g_-\sigma_x\delta\hat{\rho}(x_A(\tau), t(\tau))\big]|\Psi\rangle,
\end{eqnarray}
where $\omega_0(t)=c_0\omega_0/at$ has been taken. This choice means that a two-level detector with energy gap scaled with $c_0/at$ corresponds to a two-level detector with fixed energy gap $\omega_0$ in the time $\tau$ frame. Note that this scenario also corresponds to a two-level detector with the fixed energy gap $\omega_0$
on the $(\tau, 0)$ world line, which has been proposed to detect the timelike Unruh effect \cite{PhysRevLett.106.110404, PhysRevA.85.012306, PhysRevLett.129.160401}.

The correlation function of the density fluctuations with respect to the laboratory time $t$ along the world line $(\tau, 0)$ reads (see Supplemental Material \cite{Supplemental-Material} for details)
\begin{eqnarray}
\nonumber
\langle\delta\hat{\rho}(t(\tau))\delta\hat{\rho}(t^\prime(\tau^\prime))\rangle&=&-\frac{\rho_0}{4\pi\,mc_0}\frac{1}{(c_0\Delta\,t-i\epsilon)^2}
\\
&=&\frac{-\rho_0\,a^2}{16\pi\,mc_0^5}\frac{e^{-a(\tau+\tau^\prime)/c_0}}{\sinh^2\big[\frac{a}{2c_0}\Delta\tau-i\epsilon\big]},
\end{eqnarray}
where $\Delta\tau=\tau-\tau^\prime$ and $\langle\bullet\rangle=\langle0|\bullet|0\rangle$. With this, one can calculate the detector's response function as 
\begin{eqnarray}\label{G-omega}
\nonumber
\mathcal{G}(\omega_0)&=&g^2_-\int^\infty_{-\infty}\,d(\Delta\tau)e^{-i\omega_0\Delta\tau}e^{\frac{a}{c_0}(\tau+\tau^\prime)}
\langle\delta\hat{\rho}(\tau)\delta\hat{\rho}(\tau^\prime)\rangle
\\ 
&=&\Gamma(\omega_0)\frac{1}{e^{\frac{2\pi\omega_0c_0}{a}}-1},
\end{eqnarray}
where $\Gamma(\omega_0)=(\rho_0\omega_0g^2_-)/(2mc_0^3)$ is the spontaneous emission rate. Note that this response function is similar to that of the uniformly accelerating Unruh-DeWitt detector in $\text{R}$ region \cite{birrell1984quantum}. As such, this thermal response implies that the detector 
views the vacuum fluctuations as a thermal bath with temperature $T=|a|/2\pi$. The timelike Unruh effect is thus demonstrated.
To further explore this effect, we will explore the quantum dynamics of the impurity, and estimate the Unruh temperature with the quantum metrology approach below.

Since we are interested in the dynamics of the impurity, we trace over the degree of freedom of the analogue quantum field. Besides, 
we perform the Born approximation and the Markov approximation as a results of the weak coupling between the impurity and field.
We thus can find from Eq. \eqref{Equation-sysmtem}, the dynamics evolution of the impurity is given by the Lindblad master equation 
\begin{eqnarray}\label{master-equation}
\nonumber 
\dot{\rho}_A(\tau)=-i\Omega[\sigma_z, \rho_A(\tau)]+\mathcal{G}(\omega_0)\mathcal{L}[\sigma_+]+\mathcal{G}(-\omega_0)\mathcal{L}[\sigma_-],
\\
\end{eqnarray}
where $\mathcal{L}(O)=O\rho\,O^\dagger-O^\dagger\,O\rho-\rho\,O^\dagger\,O$. Note that $\rho_A(\tau)$ is the time-dependent state of the impurity, and $\Omega=\omega_0+\omega_L$ with $\omega_L$ being
the Lamb shift  \cite{PhysRev.72.241} as a result of the interaction with vacuum fluctuations. Specifically,  it reads 
\begin{eqnarray}
\omega_L=\frac{i}{2}[\mathcal{K}(-\omega_0)-\mathcal{K}(\omega_0)],
\end{eqnarray}
with $\mathcal{K}(\omega_0)=P\frac{1}{i\pi}\int^\infty_{-\infty}d\omega\frac{\mathcal{G}(\omega)}{\omega-\omega_0}$. Note that $P$ denotes principle value. Because of $\Gamma(\omega_0)/\omega_0\ll1$, the Lamb shift is quite small and thus usually is assumed to be negligible.

Assuming the initial state of the impurity is prepared at $|\psi(0)\rangle=\sin(\theta/2)|g\rangle+\cos(\theta/2)e^{-i\phi}|e\rangle$, one can find the solution to Eq. \eqref{master-equation} is $\rho_A(\tau)=\frac{1}{2}(\mathbf{I}+\boldsymbol{\omega}(\tau)\cdot\boldsymbol{\sigma})$ with 
\begin{eqnarray}\label{state}
\nonumber 
\omega_1&=&\sin\theta\cos(\Omega\tau+\phi)e^{-2\delta_+\tau},
\\        \nonumber 
\omega_2&=&\sin\theta\sin(\Omega\tau+\phi)e^{-2\delta_+\tau},
\\        
\omega_3&=&\cos\theta\,e^{-4\delta_+\tau}+\frac{\delta_-}{\delta_+}(1-e^{-4\delta_+\tau}),
\end{eqnarray}
where $\delta_\pm=\frac{1}{4}[\mathcal{G}(\omega_0)\pm\mathcal{G}(-\omega_0)]$, $\boldsymbol{\omega}(\tau)=(\omega_1(\tau), \omega_2(\tau), \omega_3(\tau))$, and  
$\boldsymbol{\sigma}=(\sigma_1, \sigma_2, \sigma_3)$ are Pauli matrixes.
In the long evolution time limit, $\delta_+\tau\gg1$, the impurity eventually evolves to the state 
\begin{eqnarray}
\rho_A(\infty)=\frac{e^{\beta\,H_0}}{\mathrm{Tr}[e^{\beta\,H_0}]},
\end{eqnarray}
regardless of the initial state. Here $H_0=\frac{\omega_0}{2}\sigma_z$, and $\beta=1/T$ denotes the inverse Unruh temperature. It means that the Unruh effect can be understood as a manifestation of thermalization phenomena that involves decoherence and dissipation in open quantum systems \cite{PhysRevA.70.012112}.

With the evolving state of the detector in Eq. \eqref{state}, in the following we will estimate the Unruh temperature through quantum metrology approach, similar to our previous studies \cite{QM1, QM2}.

\subsection{Optimal estimation of the Unruh effect} 
As shown above we have proved that the impurity with a specially scaled energy gap
may be exploited to observe the timelike Unruh effect in BEC. The Unruh temperature of the density fluctuations of BEC, viewed by the impurity probe, parametrizes the probe state $\rho_A(\tau)$ shown in Eq. \eqref{state}. Since the dependence of $\boldsymbol{\omega}(\tau)$
on $T$ is well understood, one can infer this temperature from the statistics of measurements 
which are made on a large ensemble of identically prepared probes. However, as a result of the random character of quantum measurement and the finite size of the ensemble, uncertainty is unavoidable in any such temperature estimate. In this regard, the theory of quantum parameter estimation plays a crucial role 
in finding the optimal measurement that minimizes this uncertainty \cite{doi:10.1142/S0219749909004839, Mehboudi_2019}.

In general, the solution of parameter estimation problem is to find an estimator which denotes a mapping $\hat{T}=\hat{T}(x_1, x_2, \dots)$ from the set 
of measurement outcomes into the space of parameters. Optimal estimators in classical estimation theory means that they can saturate the Cram\'er-Rao
inequality \cite{cramer1999mathematical, PhysRevLett.72.3439}
\begin{eqnarray}\label{CR}
V(T)\ge\frac{1}{NF(T)},
\end{eqnarray}
where $N$ is the number of measurements, $F(T)$ is the so-called Fisher Information, and $V(T)=E_T[(\hat{T}(\{x\})-T)^2]$ is the mean square error.
This inequality establishes a lower bound on the mean square error of any estimator of the parameter $T$. 
Specifically, the Fisher Information 
\begin{eqnarray}\label{FI}
F(T)=\int\,dxp(x|T)\bigg(\frac{\partial\ln\,p(x|T)}{\partial\,T}\bigg)^2.
\end{eqnarray}
Here $p(x|T)$ denotes the conditional probability of obtaining the value $x$ when the parameter has the value $T$. 
Furthermore, note that actually the mean square error is equal to the variance $\mathrm{Var}=E_T[\hat{T}^2]-E_T[\hat{T}]^2$ for unbiased estimators.

In quantum version, the conditional probability according to the Born rule reads $p(x|T)=\mathrm{Tr}[\Pi_x\rho_T]$, where $\{\Pi_x\}$ denotes a positive operator-valued measure satisfying $\int\,dx\Pi_x=\mathbf{I}$. Moreover, $\rho_T$ is the density operator parametrized by the temperature $T$ we want to estimate. 
By introducing the Symmetric Logarithmic
Derivative (SLD) $\hat{L}_T$ as the self-adjoint operator satisfying 
\begin{eqnarray}
\frac{\hat{L}_T\rho_T+\rho_T\hat{L}_T}{2}=\frac{\partial\rho_T}{\partial\,T},
\end{eqnarray}
the Fisher information \eqref{FI} then can be rewritten as 
\begin{eqnarray}\label{FI2}
F(T)=\int\,dx\frac{\Re(\mathrm{Tr}[\rho_T\Pi_T\hat{L}_T])^2}{\mathrm{Tr}[\rho_T\Pi_x]}.
\end{eqnarray} 
Therefore, for a given quantum measurement,  Eqs. \eqref{CR} and \eqref{FI2} establish the classical bound on precision, which may be achieved by a proper data processing, e.g., by maximum likelihood, which is known to provide an asymptotically efficient estimator. Since each measurement may have 
a corresponding Fisher information of its own, to find the ultimate bounds to precision one has to optimize the Fisher information over
the quantum measurements. In this regard, the Fisher information $F(T)$ of any quantum measurements is bounded by the so-called Quantum Fisher Information (QFI),
$F^Q(T)=\mathrm{Tr}[\rho_T\hat{L}^2_T]=\mathrm{Tr}[\partial_T\rho_T\hat{L}_T]$, which leads the Cram\'er-Rao bound of quantum version,
\begin{eqnarray}
\mathrm{Var}(T)\ge\frac{1}{NF(T)}\ge\frac{1}{NF^Q(T)}.
\end{eqnarray}
This inequality is valid for the variance of any estimator. 
We also define the quantum signal-to-noise ratio (QSNR) $\mathrm{Q}^2=T^2F^Q(T)$, which bounds the signal-to-noise ratio as 
$T/\Delta\,T\le\sqrt{N}\mathrm{Q}$. Hence, $\mathrm{Q}$ here can be used to quantify the ultimate sensitivity limit of our impurity thermometer.

For a qubit probe, the QFI has a simple expression in terms of the Bloch vector \cite{PhysRevA.87.022337}. Applying that to our Unruh-DeWitt detector case shown in \eqref{state}, we can obtain
\begin{eqnarray}\label{QFI}
F^Q(T)=\begin{cases}
\frac{(\boldsymbol{\omega}\cdot\partial_T\boldsymbol{\omega})^2}{1-|\boldsymbol{\omega}|^2}+|\partial_T\boldsymbol{\omega}|^2, &\omega<1, \\
|\partial_T\boldsymbol{\omega}|^2,    & \omega=1. 
\end{cases}
\end{eqnarray}
Note that the QFI for the mixed states cases (the first line of Eq. \eqref{QFI}) can be rewritten as
\begin{eqnarray}\label{QFI2}
F^Q(T)&=&\frac{(\partial_T\lambda_+)^2}{\lambda_+}+\frac{(\partial_T\lambda_-)^2}{\lambda_-},
\end{eqnarray}
where $\lambda_\pm=(1\pm|\boldsymbol{\omega}|)/2$ are two eigenvalues of the density operator of the qubit state. The corresponding eigenstates are 
\begin{eqnarray}
|p_+(\tau)\rangle&=&\sin\frac{\theta(\tau)}{2}|g\rangle+e^{-i\Omega\tau-i\phi}\cos\frac{\theta(\tau)}{2}|e\rangle,
\\
|p_-(\tau)\rangle&=&\cos\frac{\theta(\tau)}{2}|g\rangle-e^{-i\Omega\tau-i\phi}\sin\frac{\theta(\tau)}{2}|e\rangle,
\end{eqnarray}
with 
\begin{eqnarray}
\tan\frac{\theta(\tau)}{2}=\sqrt{\frac{|\boldsymbol{\omega}|-\omega_3}{|\boldsymbol{\omega}|+\omega_3}}.
\end{eqnarray}
Eq. \eqref{QFI2} means that the QFI in \eqref{QFI} consists of two terms, respectively corresponding to the Fisher information for measurements of 
$|p_+(\tau)\rangle\langle\,p_+(\tau)|$ and $|p_-(\tau)\rangle\langle\,p_-(\tau)|$. The SLD is given by 
\begin{eqnarray}\label{SLD}
\hat{L}_T=\cos\varphi|p_+(\tau)\rangle\langle\,p_+(\tau)|+\sin\varphi|p_-(\tau)\rangle\langle\,p_-(\tau)|,
\end{eqnarray}
with 
\begin{eqnarray}
\tan\varphi=\frac{\lambda_+\partial_T\lambda_-}{\lambda_-\partial_T\lambda_+}.
\end{eqnarray}
Since the SLD is directly related to the QFI and optimizes the quantum Cram\'er-Rao bound, 
measuring $\hat{L}_T$ means to minimize the uncertainty in the Unruh temperature estimate due to 
the finite number of samples. Furthermore, it is needed to point out that the SLD depends on the temperature, and thus 
some prior information on $T$ is assumed when constructing its corresponding measurement. Seen from the expression of
the SLD in Eq. \eqref{SLD}, if one wants to measure $\hat{L}_T$ in practice, then one is required to be able to efficiently 
evaluate the Bloch vector $\boldsymbol{\omega}(\tau)$ and its temperature derivatives from an accurate 
theoretical model for the detector's state $\rho_A(\tau)$. In addition, the ability to measure an arbitrary combination of $\Pi_+=\frac{\mathbf{1}+\mathbf{n}\cdot\boldsymbol{\sigma}}{2}$ and $\Pi_-=\frac{\mathbf{1}-\mathbf{n}\cdot\boldsymbol{\sigma}}{2}$ is also needed.

\begin{figure}
\centering
\includegraphics[width=0.46\textwidth]{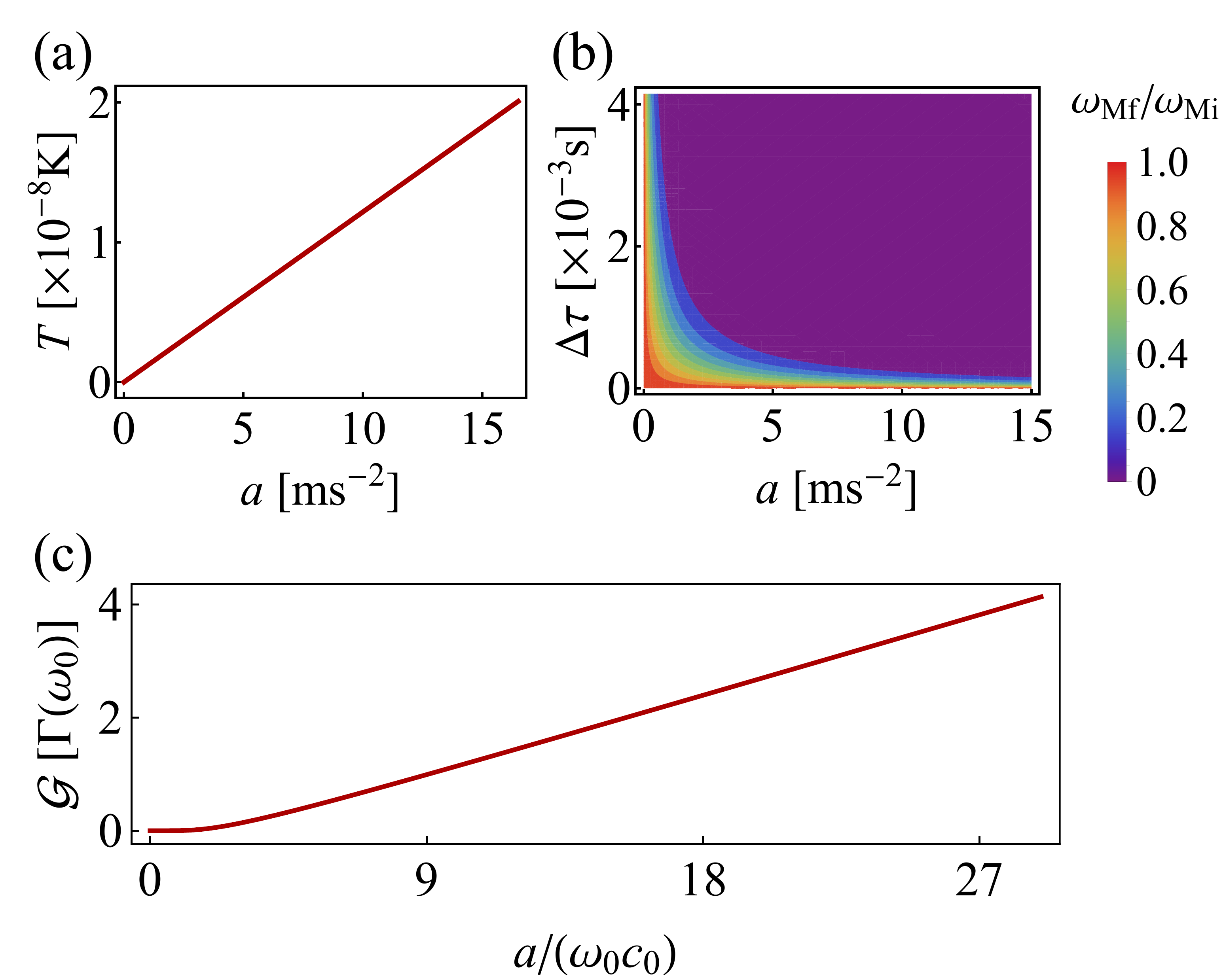}
\caption{(a) The temperature of the timelike Unruh effect via the effective acceleration; (b) the ratio of the initial, $\omega_\text{Mi}$ and final, $\omega_\text{Mf}$, frequencies in the laboratory frame as a function of the effective acceleration and the conformal evolution time of the detector; 
(c) the detector's response function (in the units of detector's spontaneous emission rate $\Gamma(\omega_0)$) as a function of the dimensionless acceleration parameter. Here we take a typical value for the speed of sound in the condensate $c_0\sim10^{-3}\mathrm{m/s}$.}\label{fig2}
\end{figure}

\begin{figure}
\centering
\includegraphics[width=0.42\textwidth]{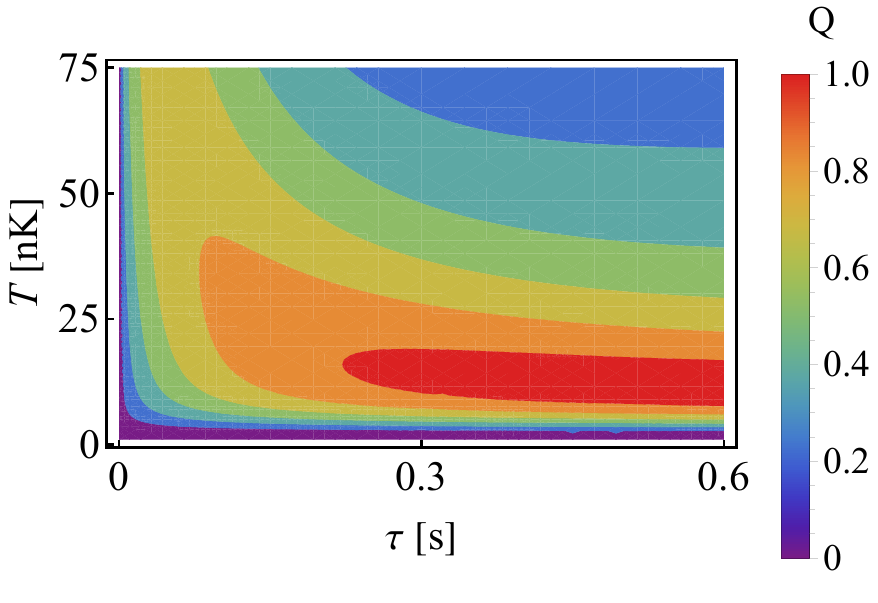}
\caption{QSNR as a function of the Unruh temperature  and the conformal evolution time of the detector for $\theta=\pi$ (i.e., the initial ground state).
Here we take a typical value for the speed of sound in the condensate $c_0\sim10^{-3}\mathrm{m/s}$, the spontaneous emission rate $\Gamma(\omega_0)/\omega_0\sim10^{-3}$, and the fixed energy gap of the detector in conformal time framework $\omega_0=2\pi\times500\mathrm{Hz}$.}\label{fig3}
\end{figure}

\section{Experimental implementation} \label{section4}
Recent experimental advances have allowed for groundbreaking observations of BEC, 
its excitation spectrum, and dynamics of impurity immersed in the BEC \cite{RevModPhys.74.875, RevModPhys.77.187, RevModPhys.91.035001}.
These relevant technologies in principle hold promise to realize our experimental scenario about timelike Unruh effect proposed above. We here make an estimate of 
the experimental parameters that are required to observe single-detector thermalization resulting from the timelike Unruh effect. 

\subsection{Unruh temperature and the conformal evolution time}

As shown above, a ``conformal observer" (i.e., in the $\tau$ framework with constant conformal frequency gap) actually will see a constant temperature
\begin{eqnarray}\label{temperature}
T=\frac{\hbar\,a}{2\pi\,c_0k_B}.
\end{eqnarray}
By taking typical value for the speed of sound in the condensate $c_0\sim10^{-3}\mathrm{m/s}$, the Unruh temperature via the effective acceleration $a$ is 
shown in Fig. \ref{fig2} (a). It is found that $1\,\mathrm{K}$ timelike Unruh temperature in the BEC approximately requires the effective acceleration as high as $10^{8}\mathrm{m/s^2}$, which is quite smaller than 
that for the massless scalar field case, on the order of $10^{20}\mathrm{m/s^2}$. This is because that the massless scalar field (or electromagnetic field) usually considered to observe the Unruh effect is replaced with the phononic field here, and the corresponding sound speed to which is much smaller than the speed of light in the vacuum. However, for the perspective of scaling the detector energy gap, we find that this scaling $a/c_0$ plays the key role in the temperature \eqref{temperature}, and $1\,\mathrm{K}$ Unruh temperature requires a scaling on the order of $100\,\mathrm{GHz}$.

We now need to clarify the correspondence between these conformal parameters and the laboratory frame parameters (i.e., Minkowski time intervals and frequencies). In particular, the relation between the time interval of observation in the laboratory frame 
and that of the conformal time $\Delta\tau$ reads 
\begin{eqnarray}\label{conformal-time}
\Delta\,t=\frac{c_0\omega_0}{a\omega_\text{Mi}}\big(e^{\frac{a\Delta\tau}{c_0}}-1\big),
\end{eqnarray}
where $\omega_\text{Mi}$ denotes the initial frequency of the two-level Unruh-DeWitt detector in the laboratory frame. Usually, since in
practice the change of the frequency is not arbitrary, the ratio of the initial, $\omega_\text{Mi}$ and final, $\omega_\text{Mf}$, frequencies in the laboratory frame
needs to be confined and clarified. This ratio is given by
\begin{eqnarray}
\frac{\omega_\text{Mf}}{\omega_\text{Mi}}=e^{-a\Delta\tau/c_0}.
\end{eqnarray}
It means the conformal time interval $\Delta\tau$ is confined. However, it is pointed out in Ref. \cite{PhysRevLett.118.045301} that unlike the electromagnetic case,  the Rabi frequency in our cold atom setup  could be tuned to zero, which gives $\omega_0=0$. Therefore, here we won't limit the ratio of the initial and final frequencies in the laboratory frame. We plot this ratio as a function of
the effective acceleration and the conformal evolution time of the detector in Fig. \ref{fig2} (b). To observe the timelike Unruh effect, the high enough effective acceleration is required to result in high enough temperature, and the long enough conformal evolution time $\Delta\tau$ of the detector is also needed to accumulate more effects on the detector (i.e., to thermalize the detector or to achieve the thermal equilibrium between 
the detector and the Unruh thermal bath). However, considering  the practical limited time interval of observation in 
the laboratory frame, we should choose proper effective acceleration and the conformal evolution time because of Eq. \eqref{conformal-time}.
We will in the following analyze the thermometric performance with considering these conditions. Furthermore, in Fig. \ref{fig2} (c) we plot the detector's response 
function (or spontaneous excitation rate) as a function of the dimensionless acceleration parameter. With the increase of the acceleration, the spontaneous excitation rate
increases monotonously. This function may be explored to verify the thermal character of the timelike Unruh effect.

 \begin{figure}
\centering
\includegraphics[width=0.42\textwidth]{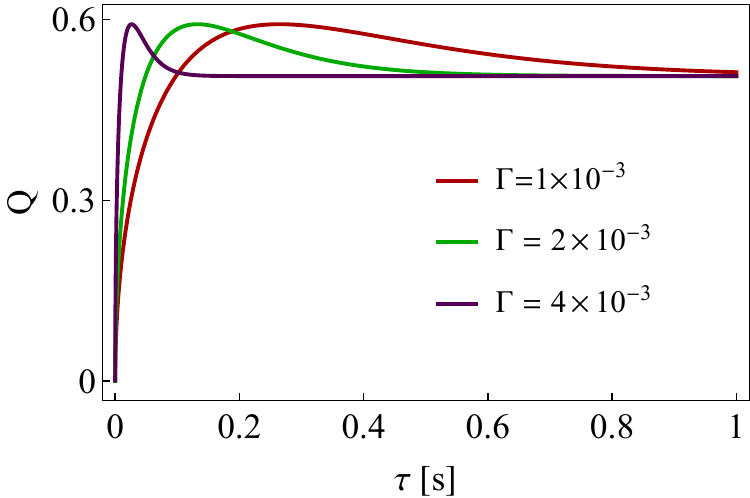}
\caption{QSNR at $T=20\,\mathrm{nK}$ as a function of the conformal evolution time of the detector for various $\Gamma=\Gamma(\omega_0)/\omega_0$ (which can be considered as the parameterized coupling strength)
by taking  $\theta=\pi$ (i.e., the initial ground state).
Here we take a typical value for the speed of sound in the condensate $c_0\sim10^{-3}\mathrm{m/s}$, and the fixed energy gap of the detector in conformal time framework $\omega_0=2\pi\times500\mathrm{Hz}$.}\label{fig4}
\end{figure}

\begin{figure*}
\centering
\includegraphics[width=0.86\textwidth]{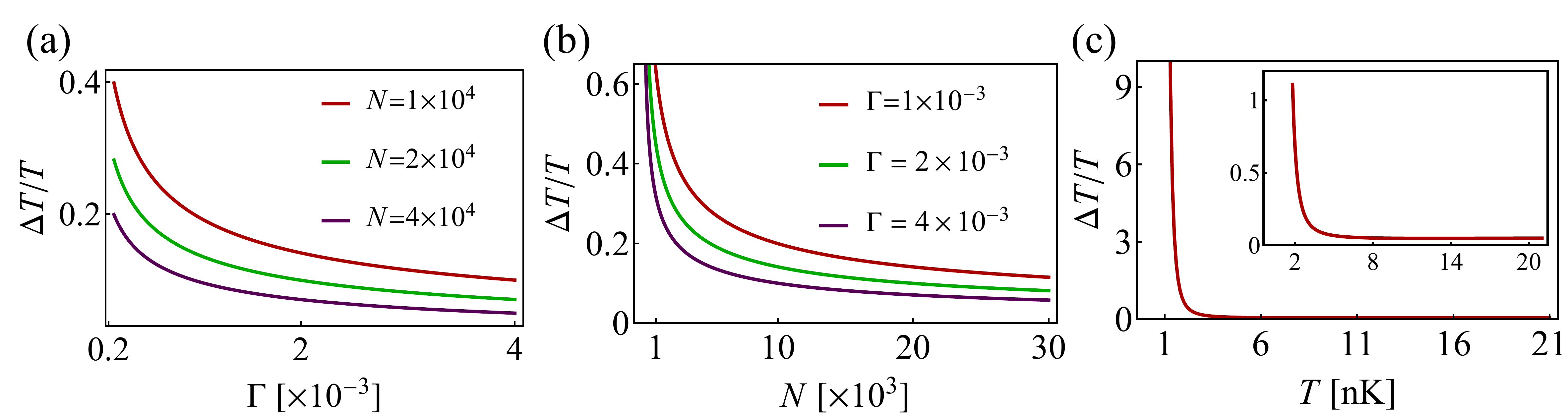}
\caption{(a) The relative error $\Delta\,T/T$ at $T=20\,\mathrm{nK}$ as a function of the spontaneous emission rate $\Gamma=\Gamma(\omega_0)/\omega_0$ (which can be considered as the parameterized coupling strength) for measurement number $N$
by taking  $\theta=\pi$ (i.e., the initial ground state); (b) the relative error $\Delta\,T/T$ at $T=20\,\mathrm{nK}$ as a function of the measurement number $N$ 
for different spontaneous emission rates $\Gamma=\Gamma(\omega_0)/\omega_0$
by taking  $\theta=\pi$ (i.e., the initial ground state); (c) the relative error $\Delta\,T/T$ at $\Gamma=4\times10^{-3}$ and $N=4\times10^4$ as a function of the experimentally feasible Unruh temperature $T$ by taking  $\theta=\pi$ (i.e., the initial ground state).
Here we take a typical value for the speed of sound in the condensate $c_0\sim10^{-3}\mathrm{m/s}$, and the fixed energy gap of the detector in conformal time framework $\omega_0=2\pi\times500\mathrm{Hz}$.}\label{fig5}
\end{figure*}

\subsection{Thermometric performance}

Note that the thermometer which measures the temperature of a BEC in the sub-$\mathrm{nK}$ regime has been investigated recently \cite{PhysRevLett.122.030403, PhysRevLett.96.130404, SR, PhysRevResearch.4.023191, PhysRevLett.125.080402}. We here consider the similar thermometer to estimate the Unruh temperature.

Fig. \ref{fig3} shows the QSNR as a function of the Unruh temperature  and the conformal evolution time of the detector for $\theta=\pi$ (i.e., the initial ground state). At a given temperature, the optimal measurement time corresponds to the maximum sensitivity, i.e., $Q_\text{max}=\max_\tau\,Q(\tau)=Q(\tau_\text{max})$. 
It would be a certain moment before the thermal equilibrium between the detector and the Unruh thermal bath, or 
 the long evolution time limit, $\delta_+\tau\gg1$, (i.e., the thermal equilibrium case), which depends on the Unruh temperature. 
 It is found that the QSNR may achieve its maximum, shown as the large red region in Fig. \ref{fig3}, in the 
 relevant temperature range which is valid within current experiments \cite{doi:10.1126/science.1088827, Nature-BEC, NP}.
 For example, if $T=20\,\mathrm{nK}$ we find $Q_\text{max}\approx0.59$, meaning that 
 an error of $\Delta\,T/T\approx10\%$ can be achieved with $N\approx280$ measurements after a time $\Delta\tau_\text{max}\approx0.265\,\mathrm{s}$.
 According to Eq. \eqref{conformal-time}, this conformal evolution time corresponds to an approximately infinite long laboratory time, i.e., 
 $\Delta\,t\rightarrow\infty$. Therefore, the maximum sensitivity $Q_\text{max}$ seems not to be achieved. However, one can still achieve the same error  
 by properly choosing the measurement time and the measurement number. For example, if $\Delta\tau=0.55\,\mathrm{ms}$, and the
 measurement number $N=50000$, one can achieve the error $\Delta\,T/T\approx10\%$ with the laboratory time $\Delta\,t\approx0.55\,\mathrm{s}$, which 
 is eminently feasible since a single gas sample may have a lifetime of several seconds \cite{PhysRevX.10.011018, PhysRevA.92.053602, PhysRevLett.105.133202}. Furthermore, instead of looking at a single impurity, we can consider, say, $1000$ to a few thousand of independent impurities \cite{PhysRevLett.121.130403, doi:10.1126/science.aaf5134}. In this case, we can effectively reduce the measurement (e.g., to $N=50$) while to achieve the same expected error $\Delta\,T/T\approx10\%$. 


The coupling strength between the detector and the field, which can be experimental controlled, plays an important role 
in the thermometric performance. In our model, the coupling strength can be embodied by the spontaneous emission rate $\Gamma(\omega_0)$ shown 
under Eq. \eqref{G-omega}. Therefore, different spontaneous emission rates could represent different coupling strength. In Fig. \ref{fig4}, we fix the Unruh temperature 
$T=20\,\mathrm{nK}$ and show the dynamical QSNR for various coupling strength. We find that the coupling strength does not affect the 
maximum sensitivity, while influences the optimal measurement time at which the maximum sensitivity can be achieved. The maximum sensitivity shifts to progressively 
later times as the coupling strength decreases. As discussed above, it seems that the maximum sensitivity can not be obtained experimentally since the optimal measurement time in conformal frame usually might correspond to infinite time in the laboratory according to Eq. \eqref{conformal-time}. 
Note that the correspondence between the conformal time and the laboratory time does not depend on the coupling strength. 
Therefore, although the experimentally feasible laboratory time ( or the conformal evolution time) is fixed and usually smaller than the optimal measurement time, by controlling the coupling strength (i.e., increasing the coupling strength) one can still in principle obtain a considerable QSNR which approaches to the maximum sensitivity. For example, if we choose the laboratory time $\Delta\,t=2\,\mathrm{s}$ which is smaller than the lifetime of a single gas sample and thus is eminently experimentally feasible, then the corresponding conformal 
evolution time $\Delta\tau\approx0.63\,\mathrm{ms}$, thus the corresponding QSNR $Q\approx0.05, 0.07, 0.1$ for $\Gamma(\omega_0)/\omega_0=10^{-3}, 2\times10^{-3}, 4\times10^{-3}$ cases, respectively.

To further understand how the coupling strength affects the estimation of the Unruh temperature, in Fig. \ref{fig5} (a) we fix an eminently experimentally feasible time 
$\Delta\tau\approx0.63\,\mathrm{ms}$ for the Unruh temperature $T=20\,\mathrm{nK}$ and show the relative error $\Delta\,T/T$ as a function of the effective coupling strength for different measurement number $N$. Remarkably, increasing the coupling strength can effectively reduce the error, meaning that the estimation of the  
Unruh temperature is more precise. For example, after 40000 measurements, the relative error for the $\Gamma(\omega_0)/\omega_0=0.2\times10^{-3}$ case
is about $22.3\%$, while for the $\Gamma(\omega_0)/\omega_0=4\times10^{-3}$ case, this achieved relative error is around $5\%$ and thus
is effectively reduced. Furthermore, as shown in Fig. \ref{fig5} (b) we can improve the estimation precision (or reduce the error) by increasing the measurements since 
$\Delta\,T/T\varpropto1/\sqrt{N}$. However, on the other hand, the huge number of measurement may take a lot of time and lower the experimental efficiency.
To achieve the same measurement error, the stronger coupling strength can effectively reduce the number of measurement under the same condition. For example,
if our target error is around $10\%$, for the $\Gamma(\omega_0)/\omega_0=1\times10^{-3}$ case, the required measurements are about $39788$, while 
for the $\Gamma(\omega_0)/\omega_0=4\times10^{-3}$ case, the required measurements are around $10028$. 
In Fig. \ref{fig5} (c) we fix an eminently experimentally feasible time $\Delta\,t=4\,\mathrm{s}$ and show the relative error $\Delta\,T/T$ 
as a function of the experimentally feasible Unruh temperature $T$ in $\mathrm{nK}$ and even sub-$\mathrm{nK}$ regime. We find that in this case 
the relative error decreases with the increase of the Unruh temperature, meaning that we can obtain more precision when estimating the Unruh temperature 
in the relatively higher temperature regime.

\section{Discussions and Conclusions} \label{section5}
The parameters were chosen above just as an example that our proposed setup is feasible, which should not be considered as the only available configuration.
Actually, we can also choose a higher effective acceleration $a$ to create a higher temperature than $\mathrm{nK}$,
Repeat the same analysis, we can get the thermometric performance, going beyond the $\mathrm{nK}$ regime.
Note that the main limited factor in our proposal is the time interval of observation in the laboratory frame since even the finite product of
the acceleration and the conformal time, i.e., $a\Delta\tau$, may corresponds to a very huge laboratory time (see Eq. \eqref{conformal-time}), which can not be accessible experimentally. However, for a fixed $a\Delta\tau$, it is possible to choose the different combinations of the acceleration and the conformal evolution time 
to improve and optimize the performance of our proposal.

Usually the Unruh effect is notoriously difficult to observe, since the temperature is so tiny for accessible values of the acceleration, $a$, namely $T=\frac{\hbar\,a}{2\pi\,ck_B}$. In other wards, to achieve an experimentally accessible Unruh temperature, quite high acceleration, which actually has been far beyond our ability, has to be required (e.g., $1\mathrm{K}$ temperature requires about $10^{20} \mathrm{m/s^2}$ acceleration). 
Here we propose to detect the so-called timelike Unruh effect \cite{PhysRevLett.106.110404, PhysRevA.85.012306, PhysRevLett.129.160401} in a BEC system. In our scenario, an inertial two-level detector, whose energy gap is continuously scaled as $1/at$ responds to the Minkowski vacuum in a manner identical to an accelerated detector with a fixed proper-energy gap. Instead of doing the real relativistic motion, e.g., quite high linear acceleration, we here just need to control the time-dependent Rabi frequency of the detector, and thus it seems to be more accessible for experiment.

Thermalization is a key feature of Unruh effect. Thus, an important question arises: is the energy gap of the detector scaled over a long enough period to allow thermalization? Let us consider a detector in conformal time is scaled between times $\tau_1$ and $\tau_2$, and assume that within the interaction time period many oscillations happens 
at the constant frequency (in the conformal time $\tau$ frame) $\omega_0$ of the detector. This requirement means  
$\Delta\tau=\tau_2-\tau_1\gg\omega_0^{-1}$. In the laboratory frame, it corresponds to $\frac{t_2}{t_1}=e^{a\Delta\tau}\gg\,e^{a/\omega_0}=e^{1/(t_1\tilde{\omega}_1)}=e^{1/(t_2\tilde{\omega}_2)}$, where $t_i$ and $\tilde{\omega}_i$ with $i=\{1, 2\}$ are respectively the laboratory time and the corresponding detector's energy gap at which. If $t_1\approx1/\tilde{\omega}_1$, then thermalization requires $t_2\gg2.71t_1$. In this regard, these thermalization conditions could lead us to 
choose certain appropriate experimental parameters to access to the optimal measurement time and thus realize 
the maximum sensitivity.

In our scheme, we estimate the Unruh temperature by monitoring the impurity atoms only, while without measuring the BEC itself destructively.
Moreover, our scheme is inherently nonequilibrium and all the underlying analysis does not assume thermalization of the impurity at the temperature of the expected
Unruh bath, thus alleviating the need for thermalization of the probe before accurate temperature estimation is feasible.

In summary, we present a concrete experimental proposal to detect the timelike Unruh effect that arises out of the entanglement between 
future and past light cones. Specifically, our model is based on an impurity with a time-dependent energy gap immersed in a BEC. We choose some typical experimentally accessible parameters to investigate 
the thermometric performance and find very low relative error of the estimated Unruh temperature can be obtained. Therefore, the preliminary estimates 
indicate that the proposed experimental implementation of the timelike Unruh effect 
is within reach of current state-of-the-art ultracold-atom experiments.

Our proposed quantum fluid platform may also allow us in the experimentally accessible regime
to explore interesting questions concerning extraction of timelike entanglement from the quantum field vacuum \cite{PhysRevA.85.012306}, Unruh effect 
induced geometric phase \cite{PhysRevLett.129.160401, PhysRevLett.107.131301}, Lorentz-invariance-violation-induced nonthermal Unruh effect \cite{PhysRevD.106.L061701}, and so on.

\begin{acknowledgments}
This work was supported by the National Natural Science Foundation of 
China under Grant No. 11905218, and the CAS Key Laboratory for Research in Galaxies and Cosmology, Chinese Academy of Science (No. 18010203). 
\end{acknowledgments}

\bibliography{Unruh-Effect-BEC}

\onecolumngrid
\vspace{1.5cm}

\newpage
\pagebreak
\clearpage
\widetext

\begin{center}
\textbf{\large Supplementary Material}
\end{center}

\setcounter{equation}{0}
\setcounter{section}{0}
\setcounter{page}{1}
\makeatletter
\renewcommand{\theequation}{S\arabic{equation}}
\section{The model}
Our detector model is inspired by the atomic quantum dot ideal originally introduced in Refs. \cite{PhysRevLett.94.040404, PhysRevLett.91.240407, PhysRevD.69.064021}. It consists of an impurity which can be consider as a two-level (1 and 2) atom, and is immersed in a one-dimensional atomic BEC at very low temperature.  The impurity is assumed to be illuminated by a monochromatic external electromagnetic field at the frequency $\omega_L$ which is close to resonance with the $1\rightarrow2$ transition $\omega_L\simeq\omega_{21}$, with a (real and positive) time-dependent Rabi frequency $\omega_0(t)$.
This kind of model has been fruitfully applied in various fields, such as investigating Casimir forces and quantum friction from Ginzburg radiation \cite{PhysRevLett.118.045301},
zero-point excitation of a circularly moving detector \cite{PhysRevResearch.2.042009}, and Lorentz-invariane-violation-induced physics \cite{PhysRevD.103.085014, PhysRevD.106.L061701}.

Specifically, the Hamiltonian of the whole system is of the form 
\begin{eqnarray}\label{Hamiltonian}
H(t)&=&H_B+H_A(t)=\int\,dk \omega_k\hat{b}^\dagger_{k}\hat{b}_{k}+\omega_{21}|2\rangle\langle2|-\bigg(\frac{\omega_0(t)}{2}e^{-i\omega_Lt}|2\rangle\langle1|+\mathrm{H.c.}\bigg)+\sum_sg_s\hat{\rho}(x_A(t))|s\rangle\langle\,s|,
\end{eqnarray}
where the last term is the collisional coupling between the impurity and Bose gas. $\hat{\rho}(x_A)=\hat{\psi}^\dagger(x_A)\hat{\psi}(x_A)$ denotes the field density operator of the atomic Bose gas, and $x_A(t)$ is the time-dependent impurity position. $g_s$ are the interaction constant 
between the impurity in $s=1, 2$ state and the condensate. In the rotating frame,
the detector's Hamiltonian including its interaction with the Bose gas can be rewritten as 
\begin{eqnarray} \label{detector-interaction}
H_A(t)&=&\omega_{21}|2\rangle\langle2|-\frac{1}{2}\omega_L(|2\rangle\langle2|-|1\rangle\langle1|)
-\frac{1}{2}\omega_0(t)(|2\rangle\langle1|+|1\rangle\langle2|)+\sum_sg_s\hat{\rho}(x_A(t))|s\rangle\langle\,s|.
\end{eqnarray}
Then, using the rotated $|g, e\rangle=(1/\sqrt{2})(|1\rangle\pm|2\rangle)$ basis and defining $g_\pm=\frac{1}{2}(g_1\pm\,g_2)$, 
we can further rewrite the Hamiltonian \eqref{detector-interaction} as
\begin{eqnarray} \label{detector-interaction2}
H_A(t)&=\frac{\omega_0(t)}{2}\sigma_z+\frac{\delta}{2}\sigma_x+\hat{\rho}(x_A(t))[g_++g_-\sigma_x],
\end{eqnarray}
where $\sigma_z$ and $\sigma_x$ are the conventional Pauli matrices, and $\delta=\omega_L-\omega_{21}$ is the detuning. 
In this rotated basis, the time-dependent Rabi frequency $\omega_0(t)$ determines the splitting between the $|g\rangle, |e\rangle$ states, while the detuning 
$\delta$ gives a coupling term.

Let us note the last term of the above Hamiltonian \eqref{detector-interaction2}
denotes the interaction between the impurity and Bose gas.
This interaction contains two terms: the first one proportional to $g_+$ is similar to the reminiscent 
coupling of a charged particle to an electric field, 
while the other resembles a standard electric-dipole coupling mediated by a coupling constant $g_-$. 
By suitably choosing the internal atomic states and properly tuning the interaction constants 
(e.g., via Feshbach resonances \cite{PhysRevLett.81.69, Inouye1998Observation}), 
the first term could be cancelled as a result of $g_+=0$ \cite{PhysRevA.74.041605, PhysRevA.77.052705, RevModPhys.82.1225}, behaving like
the analog charge neutrality. Furthermore, the atomic density operator $\hat{\rho}(x)$, as shown above, can be split into 
its average value $\rho_0$ and small fluctuations $\delta\hat{\rho}(x)$ in \eqref{density-fluctuations}. With the suitable 
detuning $\delta$ between driving frequency and the impurity's internal level space, one can 
exactly compensate the coupling to the average density, $\delta/2+g_-\rho_0=0$ \cite{PhysRevLett.118.045301, PhysRevResearch.2.042009}.
Under all these assumptions, the impurity's Hamiltonian including the interaction with the condensate 
can finally be written as 
\begin{eqnarray} \label{detector-interaction3}
H_A(t)&=\frac{\omega_0(t)}{2}\sigma_z+g_-\sigma_x\delta\hat{\rho}(x_A).
\end{eqnarray}
The coupling of the impurity to the condensate in the Hamiltonian \eqref{detector-interaction3}
 has the analogous form $g_-\sigma_x\delta\hat{\rho}(x_A)$ 
of a two-level atom dipole coupled to the one dimensional quantum scalar field at its position $x_A$. In the interaction picture with respect to the 
analogue quantum field, we can find the Hamiltonian for the whole system then can be written as 
\begin{eqnarray} \label{Interaction-Hamiltonian}
H(t)&=\frac{\omega_0(t)}{2}\sigma_z+g_-\sigma_x\delta\hat{\rho}(x_A, t).
\end{eqnarray}
Note here that $\delta\hat{\rho}(x_A, t)$ denotes the density fluctuation operator in the interaction picture.

\section{Correlation function of the density fluctuations} 
Let us present the associate expression for the correlation function of the density fluctuations. It is given by 
\begin{eqnarray}
\langle0|\delta\hat{\rho}(t, x)\delta\hat{\rho}(t^\prime, x^\prime)|0\rangle&=&\frac{\rho_0}{(2\pi)^2}\int\int\,dk\,dk^\prime(u_k+v_k)(u_{k^\prime}+v_{k^\prime})
\langle0|\big[\hat{b}_k(t)e^{ikx}+\mathrm{h.c.}\big]\big[\hat{b}_{k^\prime}(t^\prime)e^{ik^\prime\,x^\prime}+\mathrm{h.c.}\big]|0\rangle
\\  \nonumber 
&=&\frac{\rho_0}{(2\pi)^2}\int\int\,dk\,dk^\prime(u_{k}+v_{k})(u_{k^\prime}+v_{k^\prime})e^{-i\omega_k\,t+i\omega_{k^\prime}t^\prime}e^{ikx-ik^\prime\,x^\prime}(2\pi)\delta(k-k^\prime)
\\  \nonumber 
&=&\frac{\rho_0}{2\pi}\int\,dk(u_{k}+v_{k})^2e^{-i\omega_k(t-t^\prime)}e^{ik(x-x^\prime)}
\\  \nonumber 
&=&\frac{\rho_0}{2\pi}\int\,dk\frac{k^2/2m}{\omega_k}e^{-i\omega_k\Delta\,t}e^{ik\Delta\,x}
\\  
&=&-\frac{\rho_0}{4\pi\,mc_0}\frac{1}{[\Delta\,x-(c_0\Delta\,t-i\epsilon)]^2} ,
\end{eqnarray}
where $\Delta\,x=x-x^\prime$ and $\Delta\,t=t-t^\prime$ have been defined. Besides, in the last integral we have assumed the long wavelength (phononic)
regime, i.e., neglecting the quantum pressure term $\sim\nabla^2\delta\rho$, and thus we have $\omega_k\approx\,c_0k$, as done in Refs. \cite{PhysRevLett.101.110402, PhysRevLett.125.213603}.

\end{document}